\documentstyle[psfig]{elsart}
\begin{document}

\begin{frontmatter}

\title{CO adsorption on the Pt(111) surface: a comparison of a gradient
corrected functional and a hybrid functional }
\author{K. Doll}
\address{Institut f\"ur Mathematische Physik, TU Braunschweig,
Mendelssohnstra{\ss}e 3, D-38106 Braunschweig}
\maketitle

\begin{abstract}
The adsorption of CO on the Pt(111) surface
 in a $(\protect\sqrt{3} \times \protect\sqrt{3})$ pattern has been
studied with the gradient corrected functional
of Perdew and Wang and the B3LYP
hybrid functional. A slab which is periodic in two dimensions
is used to model the system.
The Perdew-Wang functional incorrectly gives the fcc site 
as the most favorable adsorption site,
in accord with a set of previous studies. The B3LYP functional
gives the top site as the preferred site. 
This confirms results from
cluster studies where it was suggested that the different splitting, 
dependent on the functional, between
highest occupied and lowest unoccupied molecular orbital, could be the
reason for this change of the adsorption site. This is supported by
an analysis
based on the projected density of states and the Mulliken population.
\end{abstract}

\end{frontmatter}


\section{Introduction}

The adsorption of CO on the Pt(111) surface is of enormous importance
in materials science, because of the high economical and 
environmental implications. This process is not at all simple and
still poses challenges to experimentalists and theoreticians.
Especially the question of the site which is occupied by the CO molecule
has triggered a huge interest in the community of those performing
ab-initio calculations. 

Experimentally, it is well established that
the top site is occupied at this coverage, and there is also a minority
of bridge sites occupied (see, e.g., reference 
\cite{Steininger1982,Blackman1988},
with low energy electron diffraction and electron-energy-loss
spectroscopy, reflection-absorption spectroscopy\cite{Hayden1983} or the
scanning tunneling microscopy experiments\cite{Stroscio1991} which
were interpreted\cite{Bocquet1996} 
in such a way that they should display CO in top and bridge sites).

First principles density functional calculations, 
however, lead to a different result: as was summarized in reference
\cite{Feibelman2001}, well converged calculations favor
the high coordination sites, and they found that the fcc (face centered
cubic) hollow site was lowest in energy. This was obtained with 
local-density or generalized gradient exchange correlation potentials,
and with various codes. 
No conclusive reason for this failure of the density functional
calculations could be given. The article initiated further studies,
and in some of them, the top site was found to be lowest in energy:
it was argued that the inclusion of relativistic effects
should lead to the correct adsorption site\cite{Orita2004,Olsen2003}. 
Others\cite{Grinberg2002} confirmed 
the wrong site preference with the top site higher
in energy, although a relativistic pseudopotential
was used. They argued that the failure was because
different bond orders were treated 
with varying accuracy due to the generalized gradient approximation.
A similar argument was that the error could be due to the incorrect
description of the HOMO-LUMO gap 
(difference between highest occupied and lowest
unoccupied molecular orbital), due to the functional employed
\cite{Gil2003,Kresse2003}, and the different occupancy of the orbitals,
depending on the adsorption site. This analysis was based on cluster 
calculations and calculations with a gradient corrected functional
for the periodic system.
The authors suggested to use hybrid functionals \cite{Gil2003}
or methods such as LDA+U \cite{Kresse2003}
to obtain a better description of the HOMO-LUMO gap.

Therefore, there are several reasons to study this system: Most important is
that the code used allows to employ hybrid functionals for periodic systems
so that the
impact of these functionals, which include Fock exchange, can be tested.
It thus makes it possible to test the suggestion \cite{Gil2003} that hybrid
functionals might lead to a preference of the top site. Note that
this suggestion was based on a cluster model, and only by an extrapolation
it could be argued that the top site would be preferred with a hybrid
functional. It is therefore interesting to perform
calculations on the periodic system and thus to remove the necessity
of a cluster approximation and the extrapolation. 
As a side effect, the approach
used here is based on local orbitals instead of plane waves, so that
a somewhat different technique is used, and a comparison with the
plane wave results for the slab calculations with the gradient
corrected functional is possible.

The article is structured as follows: after defining the computational
parameters and the slab model used, Pt bulk, the
clean Pt(111) surface, and the CO adsorption on the Pt(111) surface
are studied. This is
done with a gradient corrected and a hybrid functional, for comparison.
The focus is on the coverage of one third of a monolayer,
in a ($\sqrt 3 \times \sqrt 3$) adsorption pattern.
Finally, the results are discussed and summarized.

\section{Computational Parameters}

\subsection{Basis set and Method}

A local basis set formalism was used where the basis functions are
Gaussian type orbitals centered at the atoms as implemented in the code
CRYSTAL03 \cite{Manual03}. For Pt in the slab models, 
a scalar-relativistic pseudopotential with 18 valence electrons
was used\cite{Andrae}. The innermost $[2s3p1d]$ basis functions were
kept fixed as in the original basis set\cite{Andrae}, 
in addition one more $s$-shell
(exponent 0.59), one $sp$-shell (exponent 0.11) and one $d$-shell
(exponent 0.167) were optimized, resulting in a $[4s4p2d]$ basis set as 
a whole. For the calculations on the 
free Pt atom, which are necessary to compute the bulk cohesive energy,
the original Pt basis set was used\cite{Andrae}.
For carbon, the basis set from reference \cite{CattiMgCO3}
was chosen, with the outermost $sp$-shell (0.26) replaced by two
$sp$-shells (0.40, 0.20), i.e. $[4s3p1d]$ as a whole. 
Finally, the oxygen basis set from reference \cite{DovesiChemPhys1991}
was chosen, with outermost $sp$-exponents 0.5, 0.25, 0.1 
(instead of 0.45, 0.15) and one $d$-exponent (0.8), i.e. 
$[5s4p1d]$. The basis sets are displayed in table \ref{Basissatz}.
The numerical integration of the density functional energy was performed
on a grid, without fitting the potential with the help of an auxiliary
basis set as it had to be done with the CRYSTAL98 code.

Two functionals were used: the gradient corrected functional of
Perdew and Wang (PW91) \cite{Perdewetal}, and the hybrid functional
B3LYP.
The CRYSTAL03 code \cite{Manual03} was used for all the calculations,
except for the calculations of the free Pt atom where the Molpro2002
code \cite{Molpro2002} was used. 

\subsection{Slab model}

\begin{figure}[h]
\caption{Definition of the geometrical parameters of the slab model. 
All distances are interlayer distances.}
\label{COvonseite}
\centerline
{\psfig
{figure=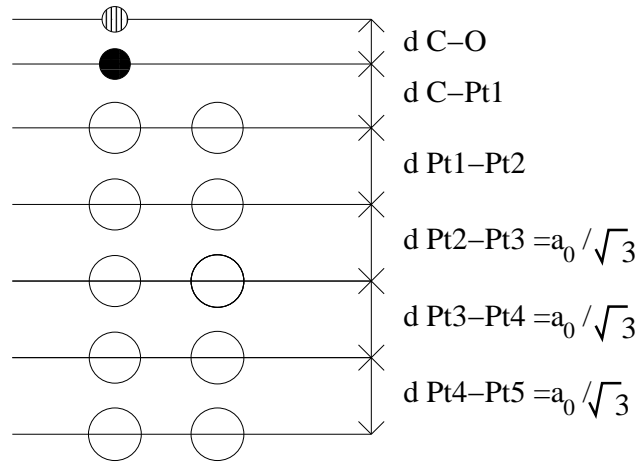,width=15cm,angle=270}}
\end{figure}

The adsorption was modeled by using a five-layer platinum slab with the
platinum atoms arranged as in the face-centered cubic lattice,
at the PW91 (and B3LYP)-optimized bulk Pt
lattice constant of 4.01 \AA \ (and 4.05 \AA).
CO was adsorbed on one side of this slab, vertical to the (111) surface. 
A supercell approach with a 
($\sqrt 3 \times \sqrt 3$) adsorption pattern as in the experiment was used.
This slab was not repeated in the third dimension so
that the model is truly two-dimensional (see figure \ref{COvonseite}).

\begin{figure}
\caption{The structures considered for CO, 
adsorbed on the Pt(111) surface, at a coverage of one third of a monolayer,
$(\protect\sqrt 3 \times \protect\sqrt 3)$R30$^\circ$ unit cell. 
The platinum atoms in the top layer 
are displayed by open circles. The considered CO adsorption sites are
the top site above the Pt atoms with number 1 (filled circles),
the threefold hollow 
sites above atoms 1,2,3 (fcc or hcp hollow, circles with horizontal lines) 
or the bridge site above atoms 2 and 3
(circles with horizontal and vertical lines). Note that the threefold hollow
sites can not be distinguished in this figure. }
\label{geometryfigure}
\centerline
{\psfig
{figure=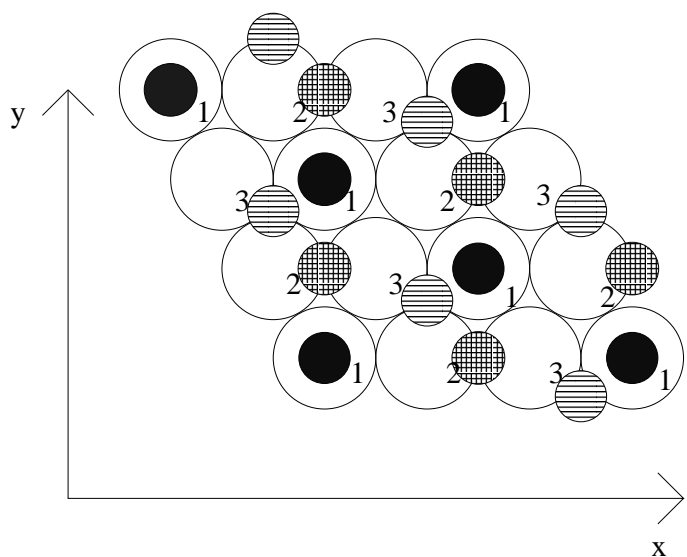,width=15cm,angle=270}}
\end{figure}

Four adsorption sites were considered (see figure
\ref{geometryfigure}): the top adsorption site with CO sitting vertically
above a platinum atom in the top layer, the bridge site with CO sitting 
above the middle of two Pt atoms in the top layer, and two
different threefold hollow sites
where the CO molecule is placed vertically above 
a platinum atom in the second (third) platinum layer.
The first of the threefold hollow sites is
referred to as the hcp (hexagonal close-packed) hollow site because
the CO molecule is vertically above  a platinum atom in the second Pt
layer.
The second of the threefold hollow sites is referred to as
the fcc hollow site because the CO molecule is vertically above third
layer Pt atoms. 

For comparison,
the vertical relaxation of the platinum atoms was simulated
in two different ways for the top site:
simulations were performed where only a uniform relaxation of the top
platinum layer was possible, and simulations where a different vertical
relaxation for the platinum atoms in the top layer was possible (i.e. 
substrate rumpling). This was done for the top site where this effect
is expected to be most important \cite{Doll2001KCu,Doll2003Ni}.
This led however, to only a very small change
in the binding energy and is thus negligible.
Also, for the hcp site, a possible second layer rumpling was considered,
i.e. the Pt atom vertically under the CO molecule in the second Pt layer
was allowed to relax. This led to an even smaller change and thus is
also considered negligible.

A $\vec k$-point sampling net of the size $16 \times 16 \times 16$ for
the bulk and $ 16 \times 16$ for the surfaces was used.
The Fermi function was smeared with a temperature of 0.01 $E_h$
(1$E_h$=27.2114 eV) to make the integration numerically more stable.

\subsection{Pt bulk and (111) surface}

The results for Pt bulk and the Pt(111) surface are summarized in tables
\ref{Ptbulktable} and \ref{Pt111table}. The lattice constant is slightly
overestimated with both functionals. 
The cohesive energy (with respect to the free atom in its $d^9s^1$ 
ground state) is slightly
underestimated with the PW91 functional, and even
more with the B3LYP functional. The fact that B3LYP underbinds can be
attributed to the admixture of Fock exchange, and the fact
that Hartree-Fock usually underbinds, especially for metals.
This was already observed, for example, in B3LYP calculations
for silver\cite{Sgroi}.

The Pt(111) surface relaxes outwards, which is different from other
metals, for example copper or silver. This is, however,
consistent with the finding
that the Mulliken charge is slightly larger in the outermost layer
(78.05 $|e|$, i.e. -0.05 $|e$). 
This is in contrast to Cu(111) and Ag(111), which relax
slightly inwards, and 
where the Mulliken charge
is slightly positive in the outermost layer (copper\cite{DollHarrison2000}: 
28.97 $|e|$, i.e. +0.03 $|e|$; 
silver\cite{DollHarrison2001}: 46.98 $|e|$, i.e. +0.02 $|e|$). 
Thus, if the atoms were considered
as spheres with a radius dependent on the charge, then the outwards
relaxation of Pt (and inwards relaxation of Cu and Ag) would be consistent
with the Mulliken charge.

\subsection{CO adsorption}

The results of the optimizations of the CO adsorption on the Pt(111) surface
are displayed in table \ref{COonPtPW91}.
First, we note that the PW91 functional gives the wrong site preference:
in agreement with most of the previous studies, the fcc hollow site is
preferred, by 0.0037 $E_h$, when comparing to the top site. 
This result is in good agreement with previous 
calculations\cite{Feibelman2001} where it was computed to be 0.10 or 0.23 eV
(0.0037 $E_h$ or 0.0085 $E_h$),
depending on the method and code used. Thus, with the scalar-relativistic
pseudopotential used in the present study, it could not be confirmed
that relativistic effects should change the site preference.
The order
of the sites is fcc, hcp, bridge, top, i.e. a higher coordination number is
favorable, however, the energy splitting is not too large. In all cases,
the outermost Pt layer relaxes slightly outwards by 0.06 - 0.08 \AA.
In the case of the top site, an additional vertical rumpling was allowed,
i.e. the Pt atom vertically below the CO molecule was allowed to relax
different than the other two Pt atoms in the top layer. 
This led, however, to only a marginal
change: instead of 0.06 \AA \ 
like the other two atoms, this third atom relaxed
outwards by 0.08 \AA. The energy was lowered by 0.0003 $E_h$, i.e. this
did not influence the site preference. Also, for the hcp site, the 
Pt atom in the second layer vertically under the CO was allowed to relax
differently (the Pt atoms in the first layer are fixed by symmetry). 
However, this also resulted in a minor change only (by 0.004 \AA, away
from CO, i.e. downwards), and the energy did virtually not change. 
Therefore, in the B3LYP study (which is more time consuming than PW91),
only a uniform relaxation of the top layer was performed.

In the following, the B3LYP results are summarized.
In the case of this hybrid functional, 
the top site has become the preferred site:
it is now lowest in energy, the fcc site is higher by 0.0016 $E_h$, the
bridge site by 0.0025 $E_h$, and the hcp site by 0.0037 $E_h$. The order
of the sites is now top, fcc, bridge, hcp, i.e. the lower coordination
sites are more favorable in B3LYP than in PW91.

It was argued that the difference between PW91 and B3LYP was the 
higher HOMO-LUMO gap which should lead to different binding energies,
depending on the site\cite{Gil2003,Kresse2003}. According to the
Blyholder model\cite{Blyholder}, 
it is thought that the CO $5\sigma$ orbital donates
charge to the Pt surface, and charge is back donated to the $2\pi^*$ orbital.
Thus the HOMO-LUMO gap, and the position of the HOMO and
LUMO with respect to the Fermi energy, are the quantity to look at. 
For the fcc site, the back donation is
larger than for the top site (\cite{Gil2003} and this work). 
The HOMO-LUMO gap is smaller in PW91, and
HOMO and LUMO are closer to the Fermi energy in PW91. 
Thus donation and
back donation is energetically more favorable with the PW91 functional,
and less favorable with the B3LYP functional where the gap is larger and
HOMO and LUMO are lower and higher with respect to the Fermi energy.
This argument was presented in reference \cite{Gil2003} and
would explain that the high coordinated
sites with a larger back donation are preferred in PW91, but not in B3LYP. 

\begin{figure}
\caption{Density of states, on CO projected, for fcc and top site, and
PW91 and B3LYP functional. 
The Fermi energy
is indicated with  dashed lines.}
\vspace{1cm}
\label{DOSfigure}
{\psfig{figure=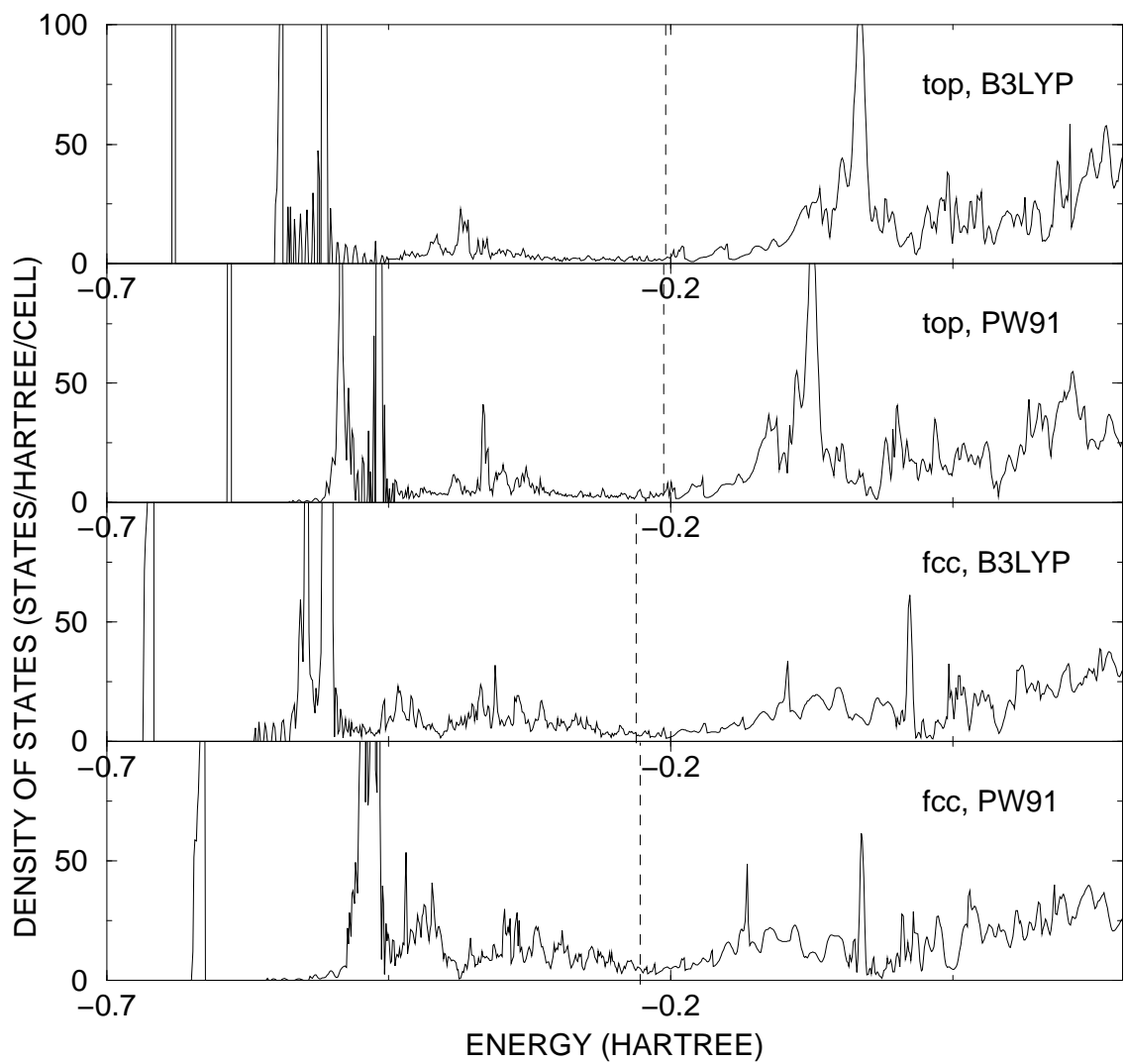,width=15cm,angle=270}}
\end{figure}

\begin{figure}
\caption{Density of states, on various orbitals projected, for the top site, 
and the PW91 functional. 
The Fermi energy
is indicated with a dashed line.}
\vspace{1cm}
\label{topPW91projectedfigure}
{\psfig{figure=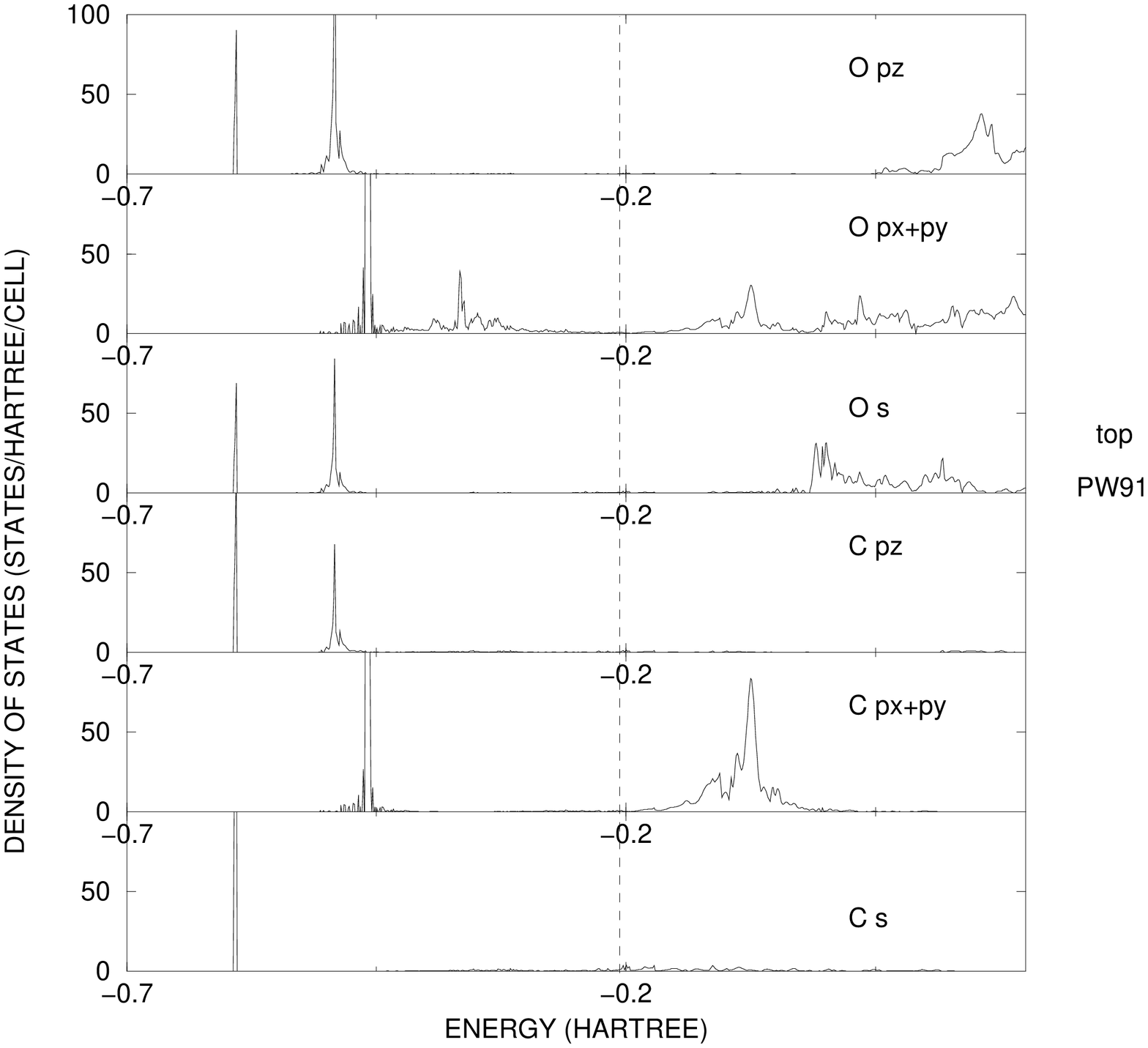,width=15cm,angle=270}}
\end{figure}

\begin{figure}
\caption{Density of states, on various orbitals projected, for the top site, 
and the B3LYP functional. 
The Fermi energy
is indicated with a dashed line.}
\vspace{1cm}
\label{topB3LYPprojectedfigure}
{\psfig{figure=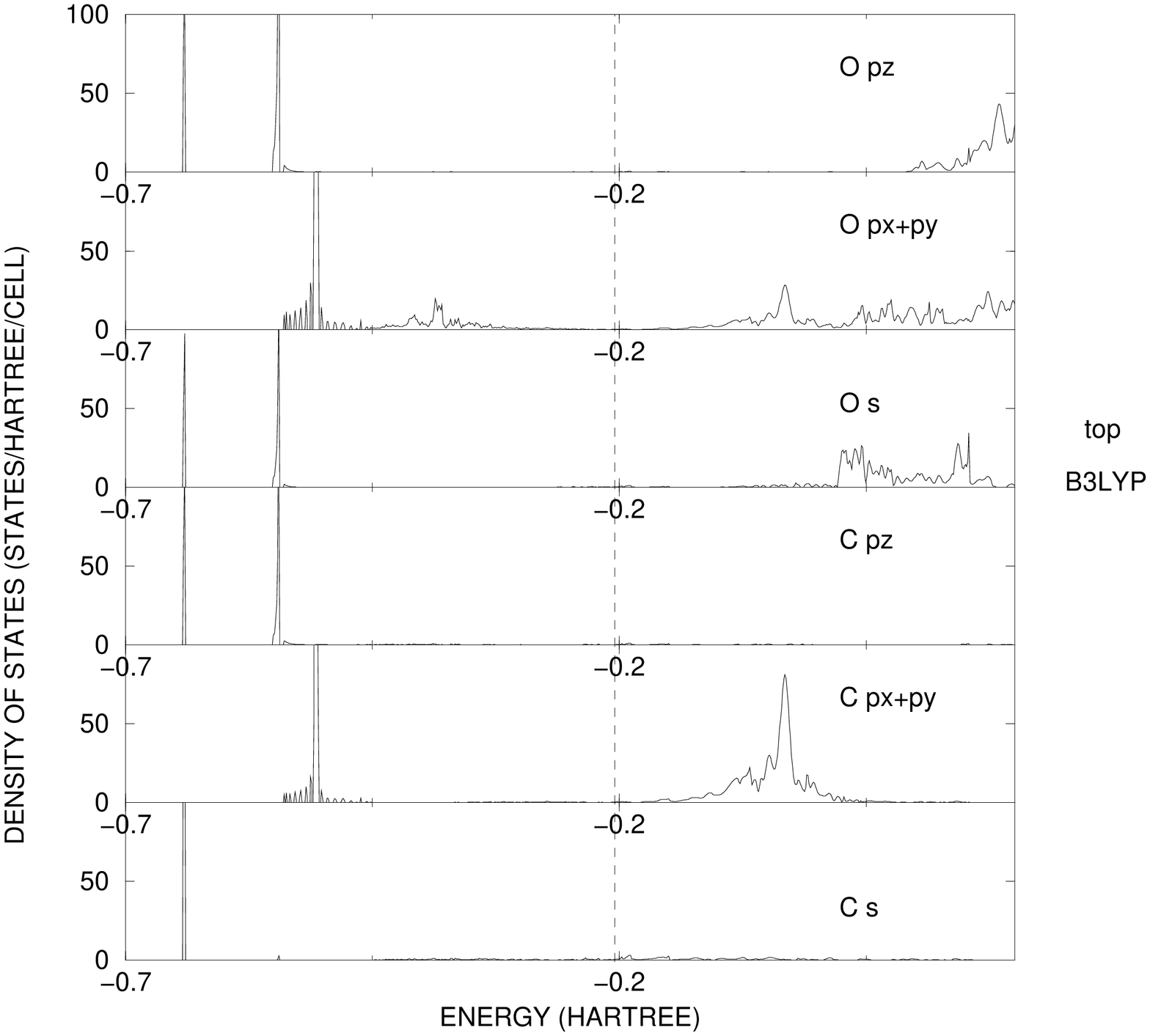,width=15cm,angle=270}}
\end{figure}

The aforementioned analysis \cite{Gil2003}
was based on a molecular system, and for the largest cluster considered
with the B3LYP functional (Pt$_{18}$-CO), the fcc site was still more
favorable than the top site, although the energy difference between fcc
and top site was smaller
than with the PW91 functional. By means of an extrapolation, using
the PW91 data as a function of the cluster size and PW91 calculations
for a slab model, 
the authors argued that in the limit of a very large cluster or a slab,
B3LYP should favor the top site. It is therefore very interesting
to study this system with a periodic code and thus 
to test this extrapolation. 
Computing the Fock exchange seems
still to be easier with codes based on Gaussian type orbitals and not
with plane wave codes, and therefore the CRYSTAL03 code is probably
the method of choice for this system. In figure \ref{DOSfigure}, the density
of states (DOS) of the CO overlayer 
is displayed for the top site and the fcc site, with PW91
and B3LYP, and the orbital projected density of states is displayed in figures
\ref{topPW91projectedfigure} and \ref{topB3LYPprojectedfigure}, for the top
site and both functionals.
The gap between the peaks in occupied and unoccupied states 
is clearly larger with the B3LYP functional, in 
agreement with the cluster study\cite{Gil2003}. 
For example, for the top site, the gap
between the peak in the DOS closest to the Fermi energy, and the first
peak above the Fermi energy is about 0.48 $E_h$ in B3LYP, but only
$0.38 E_h$ in PW91, i.e. the gap is larger by about 0.1 $E_h$
in B3LYP. 
In reference \cite{Gil2003}, the gap between $5 \sigma$ and
$2\pi^*$ orbital was determined to be 6.8 eV (0.25 $E_h$)
in PW91, and 9.5 eV (0.35 $E_h$) in B3LYP,
for an isolated CO molecule. In
the present study, the gaps are 0.42 $E_h$ in PW91, and 0.51 $E_h$
in B3LYP,
for the CO molecule adsorbed on the platinum surface, at the top site
(the $5 \sigma$ orbital
corresponds to the second peak below the Fermi energy). 
Thus the argument from the cluster studies \cite{Gil2003} can essentially be 
transferred to the periodic systems. In line with this argument is
not only the top site, but also the bridge site,
which is more favorable in B3LYP than in PW91, 
relative to the other sites. This is due to the fact that
there is less donation in the
case of the bridge site, compared to the threefold hollow sites, but more
back donation when compared to the top site.

Another interesting property to look at is the Mulliken population,
displayed in table \ref{COpopulationtable}, 
which can be interpreted as the orbital projected
densities of states, integrated up to the Fermi level (although
the calculation is in fact done in a different, much easier way).

Concerning the total charges, the main difference is that
the carbon charge is reduced for the bridge site (mainly less of
$s+p_z$ population), and even more for the top site (like bridge
less $s+p_z$, and in addition less $p_x+p_y$ population). 
As a whole, the threefold coordinated sites (fcc and hcp) have a higher
charge on CO ($\sim$ 0.3-0.4 $|e|$), 
the bridge site slightly less ($\sim$ 0.2 $|e|$), and the top site has
the lowest charge ($\sim$ 0.04-0.05 $|e|$). 

The total charge on the CO molecule is slightly larger
(by about 0.02 $|e|$)
in PW91, compared to B3LYP. This is mainly due to the sum of the
charges in the $p_x$ and $p_y$ orbitals which are for all sites
smaller in B3LYP than
in PW91. For the top site, in addition the charge in the $s$ and $p_z$ 
orbitals is larger in B3LYP than in PW91.
This is consistent with the reason
mentioned above: it is easier to put charge in the $2\pi^*$ orbital of the
CO molecule
in the case of the PW91 functional where the peaks of the unoccupied
states are closer to the Fermi energy. Also, the peaks of the
occupied states are slightly
more stabilized in B3LYP than in PW91. In the Blyholder model this means
that donation from the $5\sigma$
orbital is energetically more favorable with the PW91 functional,
and also back donation to the $2\pi^*$ orbital is more favorable with PW91.
The populations are in good agreement
with the numbers obtained with a Pt$_{18}$-CO cluster model\cite{Gil2003}.

The process of donation and back donation becomes also evident when
the orbital charges of the adsorbed CO are compared with the corresponding
charges of the free CO: the charge in $s$ and $p_z$ orbitals is reduced,
and the charge of $p_x$ and $p_y$ orbitals increases, consistent with
the Blyholder model.

We can compare the computed binding energy with the experimental
results: at zero coverage, the initial adsorption energy was measured
to be in the range from 1.04 to 1.78 eV (0.038 to 0.065 $E_h$), 
and a decrease of the adsorption energy of
the order of 0.1  - 0.2 eV (0.0037 - 0.0074 $E_h$)
was obtained for a coverage of one third
(reference \cite{Kinne2002} 
and references therein). This is in good agreement
with the data computed here, which is in the range of 0.0531 $E_h$ 
(top, B3LYP) to 0.0639 $E_h$
(fcc, PW91). The energy barrier between top and bridge site
 was experimentally measured to be in the range of 0.041 $\pm 0.007$ eV
(0.0015 $\pm 0.0003 \ E_h$) \cite{Kinne2002}
which is in reasonable agreement with the order of magnitude 
found with the B3LYP functional in this work:
the splitting between top and bridge site 
is 0.0025 $E_h$ in B3LYP, 
but 0.0013 $E_h$ with the opposite sign in PW91. 

There seems to be no experimental data for the geometry of the 
($\sqrt 3 \times \sqrt 3$) structure available. However, there is 
data from the c(4$\times$2) 
structure where also top and bridge site are
occupied\cite{Ogletree1986}. The measured bond lengths are 1.85 $\pm$ 0.01 \AA
\ for the top site and 2.08 $\pm$ 0.07 \AA \ for the bridge site.
The carbon-oxygen bond length was determined to be 1.15$\pm$0.05 \AA. 
These results are in good agreement with the results computed here
where the C-Pt bond length is 1.87 \AA \ (PW91) or 1.86 \AA \ (B3LYP) for
the top site, and 2.06 \AA \ (PW91) or 2.05 \AA \ (B3LYP) for the bridge
site; and
the computed C-O bond length is also in agreement: 1.16 (PW91)
or 1.15 \AA \ (B3LYP) for the top site, and 1.18 \AA \ (PW91) and 1.17 \AA \
(B3LYP) for the bridge site.

\section{Summary}

The adsorption of CO on the Pt(111) surface was studied with the
gradient corrected PW91 functional, and the hybrid functional B3LYP.
A scalar-relativistic pseudopotential was used for platinum in all
the studies. The adsorbate system was modeled with a slab periodic
in two dimensions.

The PW91 results confirm earlier findings\cite{Feibelman2001}: 
the fcc site is preferred
compared to the top site, which is in disagreement with the experiment
(the computed order was fcc, hcp, bridge, top; whereas in the 
experiment the top site is favored, and a minority occupancy of the
bridge site was observed).
The B3LYP results favor the top site, and the bridge site gets also
in the range of the threefold hollow sites, so that the order is
top, fcc, bridge, hcp. 

An explanation based on earlier cluster studies \cite{Gil2003}
is adopted, and supported
with the help of
the projected density of states and the Mulliken population: 
in the case of the fcc site, the $2\pi^*$ orbital has a larger occupancy
than in the case of the top site. When considering an isolated molecule,
this orbital is higher in energy relative to the highest occupied orbital,
when the B3LYP functional is used, compared to the PW91 functional.
Similarly, for the periodic system, it is found that the peak in the
density of states corresponding to this orbital is higher above the
Fermi energy in B3LYP, and the peaks corresponding to the occupied orbitals
(especially the important $5\sigma$ orbital) are lower below the
Fermi energy in B3LYP than in PW91. As a whole, PW91 will thus more favor
donation from the $5\sigma$ orbital and back donation to the $2\pi^*$
orbital (the Blyholder model). This mechanism is more important
for the threefold hollow sites, and thus PW91 favors these sites.
B3LYP, instead, does not support this mechanism so strongly which leads
to the finding that the top site is most favorable and also the bridge
site gets in the range of the threefold hollow sites. This argument, which
was initially based on cluster calculations and an extrapolation 
\cite{Gil2003}, is now confirmed with periodic calculations.
Although the
experimental result is not fully confirmed (top and minor bridge occupation),
there seems to be evidence that hybrid functionals are better able
to describe the bonding of CO to the Pt(111) surface.

\section{Acknowledgments}
All the calculations were performed at the computer centre of
the TU Braunschweig (Compaq ES 45).

\onecolumn

\newpage
\begin{table}
\begin{center}
\caption{The Gaussian basis sets used. }
\label{Basissatz}
\vspace{5mm}
\begin{tabular}{cccccc}
   &  &  exponent & contraction \\
Pt & 1$s$, 2$sp$, 3$spd$, 4$spdf$ & 
\multicolumn{2}{c}{pseudopotential \cite{Andrae}} \\
   & $5s$ &  16.5595630 &  -0.8849447\\
   &      &  13.8924400 &  1.5011228\\
   &      &   5.8536080 & -1.5529012\\
   & $6s$ &   1.2873200 &  1.0 \\
   & $7s$ &   0.59      &  1.0 \\
   & $8s$ &   0.11      &  1.0 \\
   & $5p$ &   7.9251750 &  4.9530757 \\
   &      &   7.3415380 & -5.8982100 \\
   & $6p$ &   1.9125150 &  0.3047425 \\
   &      &   1.0715450 &  0.7164894 \\
   & $7p$ &   0.4379170 &   1.0      \\
   & $8p$ &   0.11      &   1.0      \\
   & $5d$ &   3.9395310 & -0.5826439   \\
   &      &   3.5877770 &  0.5922576 \\
   &      &   1.2862310 &  0.4736921 \\
   &      &   0.5198140 &  0.5765202 \\
   & $6d$ &   0.167     &  1.0 \\ \\
 & \multicolumn{2}{c}{C}  & & \multicolumn{2}{c}{O} \\
 & exponent & contraction & exponent & contraction \\
 $1s$ &  \multicolumn{2}{c}{as in \cite{CattiMgCO3}} &
         \multicolumn{2}{c}{as in \cite{DovesiChemPhys1991}}\\ 
 $2sp$  &    3.665 & $s:$ -0.3959 $p:$ 0.2365 
         &  49.43  & $s:$   -0.00883 $p:$   0.00958 \\
        & 0.7705         & $s:$ 1.216  $p:$ 0.8606 
        &  10.47  & $s:$   -0.0915  $p:$   0.0696  \\
        & & & 3.235  &  $s:$ -0.0402     $p:$ 0.2065    \\
        & & & 1.217  &  $s:$  0.379      $p:$ 0.347    \\
  $3sp$    &   0.40  & $s:$ 1.0  $p:$  1.0 
&  0.500  & $s:$   1.0    $p:$    1.0      \\
  $4sp$    &   0.20  & $s:$ 1.0  $p:$  1.0
 &  0.250   & $s:$  1.0     $p:$   1.0   \\
  $5sp$  & & &  0.10    & $s:$  1.0     $p:$   1.0      \\
  $3d$ &   0.8  &  1.0 &  0.8     &   1.0   \\ 
\end{tabular}
\end{center}
\end{table}

\newpage
\begin{table}
\begin{center}
\caption{The ground state properties of bulk Pt. }
\label{Ptbulktable}
\vspace{5mm}
\begin{tabular}{ccccccc}
 & &  &  & \\
 & lattice constant $a_0 \ [{\rm \AA}]$ & $E_{coh} \ [E_h] $  & $B$ [GPa] & \\
PW91, this work & 4.01 & 0.192 & 227 \\
B3LYP, this work & 4.05 & 0.138 & 234 \\
Ref. \cite{Kokalj}, LDA & 3.92 & 0.276  & 310  \\
Ref. \cite{Kokalj}, PBE & 3.99 & 0.248  & 270  \\
Ref. \cite{Khein} , LDA & 3.90 & - & 307 \\
Ref. \cite{Khein} , PW91 & 3.97 & - & 246 \\
exp., Ref. \cite{Gschneider} & 3.92   &  0.215 & 278   \\
\end{tabular}
\end{center}
\end{table}

\begin{table}
\begin{center}
\caption{\label{Pt111table}The surface energy of the Pt(111) surface.}
\begin{tabular}{ccccccc}
method & relaxation of the top layer & surface energy \\
       & $[{\rm \AA}]$ & $\frac{E_h}{\rm surface \ atom}$ \\
PW91 & disallowed & 0.025 \\
PW91 & 0.05 & 0.024 \\
B3LYP & disallowed & 0.020 \\
B3LYP & 0.05  & 0.019 \\
Ref. \cite{Crljen}, PW91 &  0.023  & 0.021 \\
exp., ref. \cite{Materer} & 0.025$\pm$0.01 & \\
\end{tabular}
\end{center}
\end{table}

\newpage

\begin{table}
\begin{center}
\caption{\label{COonPtPW91} Adsorption of CO on the Pt(111) surface,
PW91 and B3LYP results. Five platinum layers are used, of which four have been
fixed at bulk positions, i.e. the distances $d_{\rm Pt2-Pt3}$,
$d_{\rm Pt3-Pt4}$, $d_{\rm Pt4-Pt5}$ are fixed at $a_{0} / 
{\protect\sqrt{3}}=2.32$ \AA \ (PW91) or 2.34 \AA (B3LYP). 
The topmost Pt layer is allowed to relax. 
$d_{\rm C-O}$ is the distance between carbon and oxygen atom, 
$d_{\rm C-Pt1}$ is
the distance between the carbon atom and top Pt layer, $d_{\rm Pt1-Pt2}$ the
distance between first and second platinum layer. 
The adsorption energy is the difference 
$E_{\rm {CO  \ at  \ Pt(111)}}-{E_{\rm Pt(111)}-E_{\rm CO}}$.}
\begin{tabular}{ccccccc}
site &   $d_{\rm {C-O}}$  & $d_{\rm {C-Pt1}}$ & $d_{Pt1-Pt2}$ &
 $E_{adsorption}$  \\
& [\AA] &    [\AA] &  [\AA] & 
$\left[\frac{E_h}{CO \ molecule}\right]$ \\
\multicolumn{5}{c}{PW91}\\
hcp hollow     &  1.19 &  1.38 &   2.40 & -0.0631 \\
fcc hollow     &  1.19 &  1.37 &   2.40 & -0.0639 \\
bridge         &  1.18 &  1.49 &   2.39 & -0.0615 \\
top            &  1.16 &  1.87 &   2.38 & -0.0602 \\
\multicolumn{5}{c}{B3LYP}\\
hcp hollow     &  1.18 &  1.35 &   2.45 & -0.0494 \\
fcc hollow     &  1.18 &  1.33 &   2.43 & -0.0515 \\
bridge         &  1.17 &  1.46 &   2.43 & -0.0506 \\
top            &  1.15 &  1.86 &   2.39 & -0.0531 \\
\end{tabular}
\end{center}
\end{table}

\newpage
\begin{table}
\begin{center}
\caption{Orbital-projected charge of CO on different adsorption sites.
Note that the charge of the $d$ basis functions is essentially constant
($\sim$ 0.1 $|e|$ for the molecule) and not displayed }
\label{COpopulationtable}
\begin{tabular}{cccccccc}
site & \multicolumn{6}{c}{charge, in $|e|$} \\
& C $s+p_z$ & C $p_x+p_y$ & O $s+p_z$ & O $p_x+p_y$  & C total & O total &
CO total \\
 & \multicolumn{6}{c}{PW91}\\
hcp hollow   & 4.489 & 1.618 & 5.217 & 2.916 & 6.187 & 8.167 & 14.354 \\
fcc hollow   & 4.503 & 1.624 & 5.218 & 2.914 & 6.207 & 8.165 & 14.372 \\
bridge       & 4.374 & 1.590 & 5.229 & 2.914 & 6.041 & 8.178 & 14.219 \\
top          & 4.366 & 1.394 & 5.235 & 2.946 & 5.833 & 8.220 & 14.053 \\
 & \multicolumn{6}{c}{B3LYP}\\
hcp hollow   & 4.490 & 1.568 & 5.210 & 2.946 & 6.141 & 8.191 & 14.332 \\
fcc hollow   & 4.506 & 1.576 & 5.209 & 2.944 & 6.168 & 8.189 & 14.357 \\
bridge       & 4.379 & 1.532 & 5.224 & 2.942 & 5.995 & 8.202 & 14.197 \\
top          & 4.387 & 1.334 & 5.235 & 2.968 & 5.798 & 8.241 & 14.039 \\
 & \multicolumn{6}{c}{free CO molecule, PW91 ($r_e$=1.15 \AA)} \\
& 4.553 & 1.018       & 5.412 & 2.922 & 5.631  & 8.369 & 14.000  \\
 & \multicolumn{6}{c}{free CO molecule, B3LYP ($r_e$=1.14 \AA)} \\
& 4.551 & 0.988       & 5.415 & 2.946 & 5.603  & 8.397 & 14.000 \\
\end{tabular}
\end{center}
\end{table}

\clearpage

\newpage

\newpage

\newpage

\newpage

\newpage

\end{document}